# How to find molecules with long-lasting charge migration?

Alan Scheidegger*, Nikolay V. Golubev* and Jiří Vaníček*

*Abstract:* Under certain conditions, the ionization of a molecule may create a superposition of electronic states, leading to ultrafast electron dynamics. If controlled, this motion could be used in attochemistry applications, but it has been shown that the decoherence induced by the nuclear motion typically happens in just a few femtoseconds. We recently developed an efficient algorithm for finding molecules exhibiting long-lasting electronic coherence and charge migration across the molecular structure after valence ionization. Here, we first explain why the but-3-ynal molecule is a promising candidate to study this type of ultrafast electron dynamics. Then, we use the 3-oxopropanenitrile molecule, which does not show long-lasting charge migration in any of three different ionization scenarios, as an example demonstrating that several different properties must be fulfilled simultaneously to make the attochemistry applications possible.

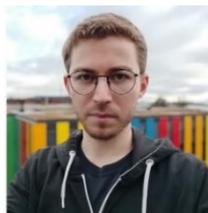

**Alan Scheidegger** finished his Master of Science in Molecular and Biological Chemistry at EPFL in 2020. During this time, he spent a semester in Prof. Clémence Corminboeuf's group studying photoswitches and conducted a master thesis researching electronic coherence following ionization in molecular systems in the group of Prof. Jiří Vaníček. Alan Scheidegger then started his PhD in the same group to develop efficient mixed quantum-semiclassical methods to perform nonadiabatic dynamics.

## 1. Introduction

Femtochemistry, recognized by the attribution of the Nobel prize to Ahmed Zewail in 1999, made it possible to study and control the rearrangement of nuclei in a molecule with laser pulses. Pushing the limit of time resolution even further, attochemistry aims to control the chemical reactivity of molecules by directly manipulating the electron motion on its natural time scale[1] with applications including selective bond breaking[2], bond formation[3] or modifying the relaxation time after photoexcitation.[4] To achieve this goal, it is crucial to (I) prepare the desired electronic coherences, (II) have an appropriate observable property that can be manipulated by controlling the electron motion, and (III) win the race against various decoherence effects, which could destroy the electron dynamics. The pioneering work of Weinkauf, Schlag and co-workers[5] showed that local ionization of peptide chains at the C-terminal side and subsequent UV photofragmentation led to the breaking of the molecules at the N-terminal end, due to the migration of the positive charge after the ionizing pulse. This finding motivated numerous experimental[2,6] and theoretical[7,8] studies, which demonstrated that the ionization of a molecule by an ultrashort laser pulse is a convenient process to test the paradigm of attochemistry. Under certain conditions, the ionization of a molecule may bring the system to a coherent superposition of electronic states. The created positive charge, or hole, will oscillate across the molecular structure with a period inversely proportional to the energy gap between the populated cationic states. This introduces a time dependence of the chemical reactivity, exploitable by a second pulse, which would generate different outcomes depending on the delay. The hole dynamics is purely driven by electron correlation and has been called "charge migration"[8,9] to differentiate it from charge transfer[10,11], which is caused by the nuclear motion. However, the nuclear and electronic dynamics are coupled since the molecular geometry determines the electronic structure and, as a result, the forces acting on the nuclei. It has been shown that typically the nuclear motion will stop the hole oscillations within a few femtoseconds.[12,13] Several theoretical studies were conducted to understand the influence of the molecular structure on the hole dynamics.[14,15] In these studies, most of the simulations were performed in the absence of nuclear motion and their results may, therefore, be very challenging or even impossible to observe experimentally if the decoherence is fast. If, on the contrary, the electronic coherence persists long enough, it has been proposed[16] that attosecond transient absorption spectroscopy could be used to probe the hole motion in real time.

*Correspondence: Prof. J. Vaníček, E-mail: jiri.vanicek@epfl.ch, Laboratory of Theoretical Physical Chemistry, Institut des Sciences et Ingénierie Chimiques, Ecole Polytechnique Fédérale de Lausanne (EPFL), Av. F.-A. Forel 2, CH-1015 Lausanne, Switzerland,
A. Scheidegger, E-mail: alan.scheidegger@epfl.ch
Prof. N. Golubev, E-mail: ngolubev@arizona.edu, Department of Physics, University of Arizona, Tucson, AZ 85721, USA.

We recently presented an efficient algorithm for finding suitable candidates for experimental studies of long-lasting charge migration and applied it to a large variety of small organic molecules.[17] Our algorithm, based on a semiclassical analysis of the electronic coherence, consists of several steps, starting from the construction of the neutral ground state geometry to the evaluation of the coupled electron-nuclear dynamics. In this process, we require the presence of a hole-mixing structure in the ionization spectrum, long-lasting electronic coherence and a few oscillations of the created charge along the molecular structure to fulfill the three necessary properties for attochemistry applications. After reviewing the algorithm in Sec. 2, we present the but-3-ynal molecule as an example of a promising candidate for an experimental study of ultrafast charge migration. In contrast, the 3-oxopropanenitrile molecule is used to illustrate how the absence of any one of the three necessary conditions prevent long-lasting hole dynamics.

## 2. Theoretical framework

### 2.1 Ionization and hole-mixing

Treating the ionization explicitly is challenging as it requires describing the departing electron and its effect on those left behind.[18] A convenient approach starts from the sudden ionization approximation, where the initial cationic state $|\Psi^{N-1}(0)\rangle$, with $N-1$ electrons, is determined by the projection of the neutral vibrational and electronic ground state onto the ionic subspace. To describe it mathematically, we first expand the $I$-th cationic eigenstate $|\Psi_I^{N-1}\rangle$ in the second-order configurational interaction series:

$$|\Psi_I^{N-1}\rangle = \sum_k c_k^I \hat{c}_k |\Phi_0^N\rangle + \sum_{k<l}\sum_a c_{kla}^I \hat{c}_a^\dagger \hat{c}_k \hat{c}_l |\Phi_0^N\rangle,$$

(1)

where $\hat{c}_a^\dagger$ is the creation operator adding an electron to the the $a$-th orbital of the neutral electronic state $|\Phi_0^N\rangle$ and $\hat{c}_k$ is the annihilation operator removing an electron from the $k$-th orbital. The first term includes the one-hole (1h) configurations, while the second term represents the two-holes-one-particle (2h1p) contributions, where, in addition to the removal of an electron, another one is promoted to a virtual orbital. The amplitudes of the configurations $c_k^I$ and $c_{kla}^I$ can be computed with electronic structure methods, such as the algebraic diagrammatic construction (ADC)[19,20] and the equation-of-motion coupled-cluster method for ionization potential (EOM-CC-IP).[21–23] The hole-mixing phenomenon appears when the same 1h configuration contributes to two different cationic eigenstates. In this case, the removal of the electron associated to this configuration will lead to the population of correlated ionic states and, hence, to the creation of a superposition

$$|\Psi^{N-1}(0)\rangle = \sum_I a_I |\Psi_I^{N-1}\rangle$$

(2)

of states, where the probability amplitude $a_I$ can be determined from the 1h configuration amplitudes $c_k^I$.[17]

### 2.2 Semiclassical nuclear dynamics

The wavepacket propagation on each populated cationic state is performed with the thawed Gaussian approximation (TGA).[24] In this framework, the nonadiabatic couplings are neglected and a single Gaussian wavepacket

$$\chi_I(\mathbf{R}, t) = \sqrt{p_I} \exp\left\{\frac{i}{\hbar}\left[\frac{1}{2}\left(\mathbf{R} - \mathbf{R}_t^I\right)^T \cdot \mathbf{A}_t^I \cdot \left(\mathbf{R} - \mathbf{R}_t^I\right) + \left(\mathbf{P}_t^I\right)^T \cdot \left(\mathbf{R} - \mathbf{R}_t^I\right) + \gamma_t^I\right]\right\},$$

(3)

is propagated on each populated cationic state $I$. Here $\mathbf{R}_t^I$ and $\mathbf{P}_t^I$ denote the position and momentum, $\mathbf{A}_t^I$ is a complex symmetric width matrix with a positive definite imaginary part, and $\gamma_t^I$ is a complex number, whose real part contains the phase while the imaginary part guarantees the normalization. The population $p_I$ defines the fraction of the total molecular wavefunction present in the electronic state $I$. Since the nonadiabatic effects are not taken into account in the TGA, the populations $p_I$, which can be identified by the sudden ionization model, remain constant in time.

The initial position and the width of the wave packet are determined from the local harmonic approximation to the ground electronic surface at the equilibrium geometry. The initial momentum and phase are assumed to be zero at the beginning of the propagation.

While the position and momentum simply follow classical Hamilton's equations of motion, the phase and width are propagated with the single-Hessian variant[25] of the TGA, with the reference Hessian of each cationic state chosen at the Franck-Condon point.

Electronic coherence $\chi_{IJ}$ between two cationic states $I$ and $J$ is given by the overlap

$$\chi_{IJ}(t) = \sqrt{p_I p_J} \int \chi_I^*(\boldsymbol{R},t)\chi_J(\boldsymbol{R},t)d\boldsymbol{R} \qquad (4)$$

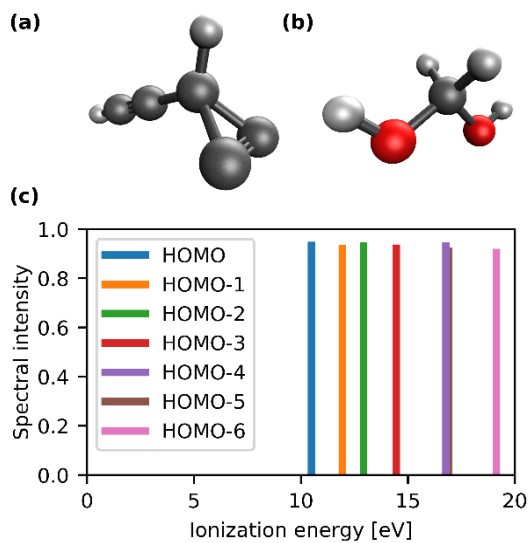

Fig. 1. (a) Geometry of a molecule failing the geometry optimization. (b) Geometry of a molecule with no hole-mixing in its (c) ionization spectrum.

of the corresponding wavepackets. Initially equal to $\sqrt{p_I p_J}$, the overlap $\chi_{IJ}(t)$ tends to zero as the wavepackets on the different electronic surfaces start moving in different directions. Although it is one of the simplest semiclassical methods, the TGA is exact for infinitesimally short times and has been shown to be accurate on ultrashort time scales of interest here. In addition, the TGA can provide insight into the decoherence mechanism.[26]

### 2.3 Filtering

Our algorithm takes molecules in the SMILES format, which describes connectivity between the atoms with a chain of characters. This allows us to quickly construct a new dataset or modify an existing one by easily switching or adding chemical groups. Using OpenBabel 3.0.0,[27] the 3D geometry is first constructed with the MMFF94 force field[28] and then optimized at the wB97XD/6-311++G(d,p) level.[29] The ionization spectrum is computed with the ADC(3) method where the non-correlated Hartree-Fock reference is obtained from the GAMESS UK 7.0 electronic structure package.[30] At this point, we check that a sufficient hole-mixing between the first few cationic states is present to ensure the possibility to create a superposition of states by ionization. At the same time, we make sure that the cationic states of interest are sufficiently separated from the neighboring ones, in order to be able to neglect nonadiabatic effects during the wavepacket propagation. If those conditions are fulfilled, we perform the classical molecular dynamics using the time-dependent density-functional theory (TDDFT)[31,32] at the wB97XD/6-311++G(d,p) level using the Verlet algorithm and a timestep of 0.25 fs for a total duration of 25 fs. Afterwards, the TGA is applied to the classical trajectories, and the time-dependent overlap of Gaussian wavepackets propagated on the two surfaces yields the time evolution of the electronic coherence. For molecules with long-lasting electronic coherence, the propagation is repeated using the EOM-CC-IP method and DZP basis set using the Q-Chem package[33] to verify the TDDFT results.

### 3. Examples of studied molecules

Our filtering from Sec. 2.3 acts as a funnel that selects only the molecules compatible with our theoretical approach. We observed that only a fraction of all considered molecules satisfied our requirements and passed all steps of the algorithm. Of the 253 molecules we scanned, 230 were successfully optimized in the neutral ground state. The reason for the failed optimizations of 23 molecules is that PubChem contains a great variety of chemical compounds and, as a consequence, also includes unstable molecules, such as the 3-ethynylcyclopropyne molecule, shown in Fig 1(a). In this case, the 3-atom ring induces a high strain, leading to the breaking of the cycle during the optimization. The most selective step comes after the evaluation of the ionization spectrum, because several conditions must be fulfilled simultaneously before one can proceed with the semiclassical nuclear dynamics. As an example, we show the structure of the methanediol molecule and its spectrum in panels (b) and (c) of Fig. 1. All cationic states up to 20 eV are predominantly formed by single 1h configurations generated by the ionization of specific orbitals, and no hole-mixing is present.
Of the 230 stable structures, only 29 molecules had both strong hole-mixing and well separated cationic states. In the last step, we applied the semiclassical dynamics propagation and evaluation of the electronic coherence to these remaining 29 candidates.

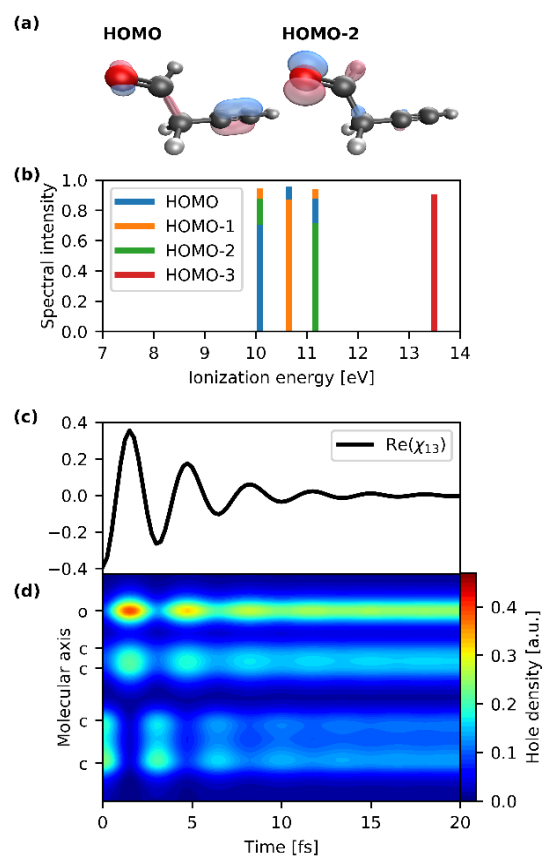

Fig. 2. Ionization spectrum and the coupled electron-nuclear dynamics triggered by the ionization out of the HOMO of the but-3-ynal molecule. (a) HF molecular orbitals involved in the hole-mixing. (b) First four cationic states. (c) Time evolution of the electronic coherence between the first and the third cationic states. (d) Time evolution of the hole density along the molecular axis. Scheidegger et al., J. Chem. Phys. **2022**, 156, 034104; licensed under a Creative Commons Attribution (CC BY) license.[17]

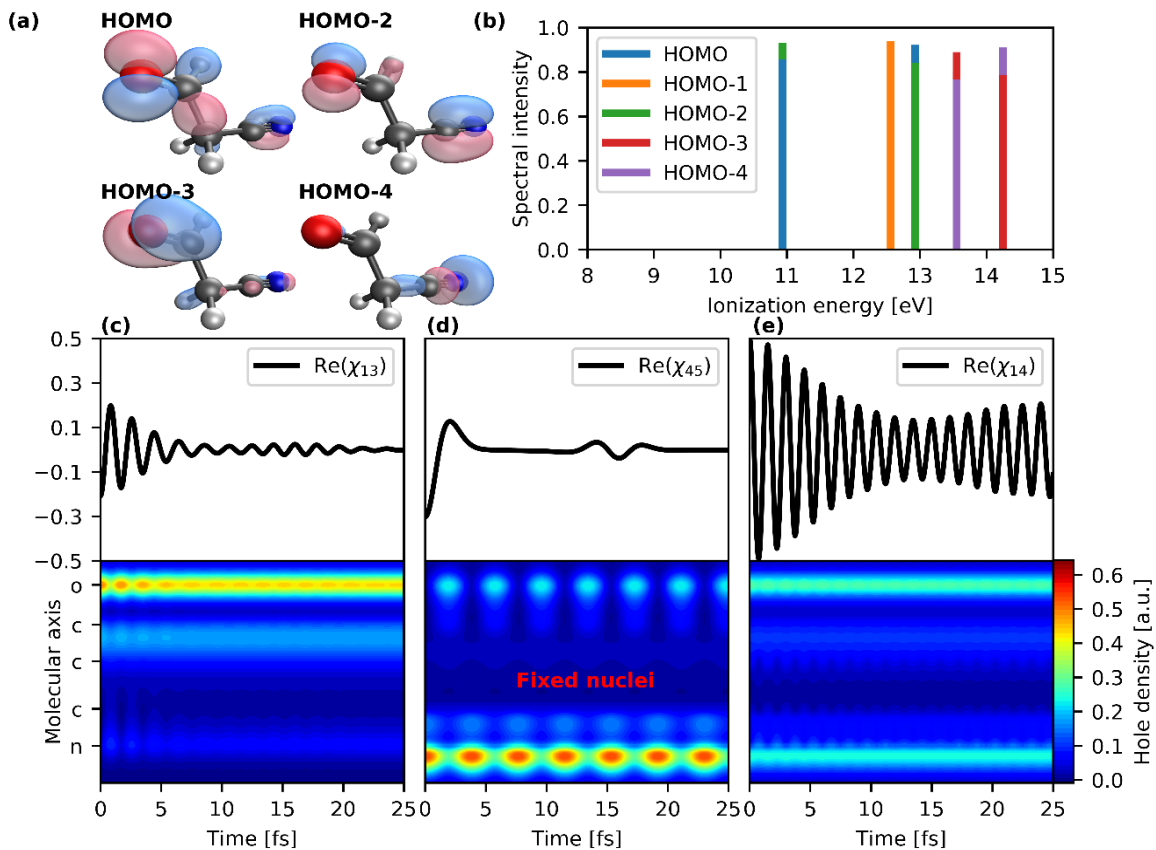

Fig. 3. Ionization spectrum and the coupled electron-nuclear dynamics triggered by the creation of a cationic superposition of states in the 3-oxopropanenitrile molecule. (a) HF molecular orbitals. (b) First five cationic states. (c-e) Time evolution of the electronic coherence and charge migration after the creation of a superposition of the (c) first and third, (d) fourth and fifth and (e) first and fourth cationic states.

### 3.1 but-3-ynal
One of the most promising molecules that we have found is the but-3-ynal molecule [Fig. 2(a)]. As shown in Fig. 2(b), strong electron correlation mixes the HOMO and HOMO-2 in the first and third cationic states. We simulated the ionization out of the HOMO and neglected its small contribution in the second spectral line. In this case, the electronic coherence, shown in Fig. 2(c), persists for the first 10 fs and at least 3 clear oscillations are observed. The two relevant orbitals are strongly localized at the opposite ends of the molecule, leading to a clear migration of the hole across the molecular structure.

### 3.2 3-oxopropanenitrile
The 3-oxopropanenitrile molecule is structurally similar to the but-3-ynal but contains a cyanide group instead of the carbon-carbon triple bond [Fig. 3(a)]. The first and third cationic eigenstates, shown in Fig. 3(b), consist of a mixture of the HOMO and HOMO-2, similarly to the but-3-ynal. In addition, the fourth and fifth cationic states also present mixing, but between the HOMO-3 and HOMO-4. The two energy gaps are respectively 2.00 eV and 0.69 eV, leading to the possibility to induce electron dynamics with a period of 2.1 fs or 6.0 fs. Both ionization scenarios respect the conditions set in our algorithm and explain why this molecule has been selected.
Below, we use three different ionization scenarios to illustrate how the lack of charge migration, long-lasting electronic coherence, or hole-mixing prevent the molecule to be a suitable candidate for experimentation.

In the first case, shown in Fig. 3(c), we simulated the ionization out of the HOMO. The hole-mixing requirement is fulfilled and the electronic coherence lasts for about 10 fs, with a few clear oscillations. A slight revival of the coherence occurs around 15 fs. However, the hole dynamics is almost imperceptible. This is mostly due to the weakness of the hole-mixing and partially due to the insufficient difference between the densities of the two involved orbitals.

In Fig. 3(d), we show the results following the ionization of the HOMO-4 which is also involved in hole-mixing. In this case, the decoherence happens faster than the oscillation period and the hole would not have time to perform a single oscillation. Consequently, we simulated the charge migration without accounting for nuclear motion. The HOMO-4 is strongly localized on the nitrogen while the HOMO-3 is centered around the CO bond. This explains the clear oscillations from one end of the molecule to the other. The period of oscillation is shorter than expected from the ionization spectrum due to slightly different energy gaps predicted by ADC(3) and EOM-CC-IP.

Finally, we also computed the electronic coherence between the first and fourth cationic states, as shown in Fig. 3(e), despite the absence of hole-mixing. Since the sudden ionization does not lead to the creation of the electronic wavepacket in this case, we artificially created an equally weighted superposition of the corresponding states. Surprisingly, the electronic coherence does not vanish completely. Instead, one can see that the coherence starts to increase again at around 15 fs. This phenomenon of the revival of the electronic coherence has already been observed in silane but on a longer time scale.[34] Unfortunately, without electron correlation the charge migration is barely perceptible. Despite not being suited for charge migration measurements, the 3-oxopropanenitrile molecule shows that the electronic coherence is not doomed to quickly vanish due to the nuclear motion but can persist for a few dozen femtoseconds.

### 4. Conclusion and outlook

Using a semiclassical description of the nuclear dynamics, we simulated the charge migration of the but-3-ynal and 3-oxopropanenitrile molecules after ionization. While the former is an excellent candidate for further experimentation, the latter illustrates how the lack of hole dynamics, long-lasting electronic coherence, or hole mixing prevents the observation of ultrafast charge migration following ionization. Because the field of attochemistry is still in its early stage, we hope that our method[17] will help in the discovery of new charge-directed chemical reactions. The semiclassical propagation of the system is the most computationally expensive step of our algorithm. An important efficiency improvement could be achieved by adding a preselection filter to remove molecules with very fast decoherence. This could be implemented by first performing the propagation using the global harmonic approximation, which only requires the energies, gradients, and Hessians of the potential energy surfaces at the equilibrium geometry.

Selecting molecules, for which we can neglect nonadiabatic effect, is quite restricting. An improvement would be achieved by performing nonadiabatic nuclear wavepacket propagation with one of the variants of the surface-hopping method including decoherence correction, such as that developed by Granucci et al.[35]

*Acknowledgements*

A.S. is grateful to Metrohm and the Swiss Chemical Society for the award for the best oral presentation in computational chemistry. The authors acknowledge financial support from the Swiss National Science Foundation through the National Center of Competence in Research MUST (Molecular Ultrafast Science and Technology) (phase III) (51NF40-183615) and from the European Research Council (ERC) under the European Union's Horizon 2020 Research and Innovation Programme (Grant Agreement No. 683069–MOLEQULE). N.V.G. acknowledges support by the Branco Weiss Fellowship—Society in Science, administered by the ETH Zürich.

Received: xx.xx.2023